%% file: asilomar23_paper.tex
\documentclass[conference]{IEEEtran}
\IEEEoverridecommandlockouts
\usepackage{cite}
\usepackage{amsmath,amssymb,amsfonts}
\usepackage{graphicx}
\usepackage{textcomp}
\usepackage{xcolor}
\usepackage[font=footnotesize, labelsep=none]{caption}
\usepackage{subcaption}
\usepackage[hidelinks]{hyperref}
\usepackage{mathtools}
\usepackage{multirow}

\usepackage[
	all=normal,
	leading=tight,
	leadingfraction=0.98, 
	]{savetrees}

\def\BibTeX{{\rm B\kern-.05em{\sc i\kern-.025em b}\kern-.08em
		T\kern-.1667em\lower.7ex\hbox{E}\kern-.125emX}}

\newcommand{\shrt}[1]{\hspace*{1.5pt}{#1}\hspace*{1.5pt}}
\newcommand{\?}{\hspace*{-0.5pt}}

\newcommand{\m}[1]{\mathbf{#1}} 

\newcommand{\lt}[1]{_{\text{#1}}}

\DeclareMathOperator{\Ex}{\mathbb{E}}

\begin{document}
	
\title{Head Orientation Estimation with Distributed Microphones Using Speech Radiation Patterns
	\thanks{
		This project has received funding from the SOUNDS European Training Network -- an European Union’s Horizon 2020 research and innovation programme under the Marie Skłodowska-Curie grant agreement No. 956369.}
}

\author{
	\IEEEauthorblockN{
		Kaspar M\"uller\IEEEauthorrefmark{1},
		Bilgesu \c{C}akmak\IEEEauthorrefmark{2},
		Paul Didier\IEEEauthorrefmark{2},
		Simon Doclo\IEEEauthorrefmark{3},
		Jan \O{}stergaard\IEEEauthorrefmark{4},	
		Tobias Wolff\IEEEauthorrefmark{1}
	}
	\IEEEauthorblockA{
		\IEEEauthorrefmark{1}%
		Cerence GmbH, Acoustic Speech Enhancement, Ulm, Germany,
		Email: \{kaspar.mueller, tobias.wolff\}@cerence.com}
	\IEEEauthorblockA{
		\IEEEauthorrefmark{2}%
		Department of Electrical Engineering (ESAT),
		KU Leuven,
		Leuven, Belgium \\
	}
	\IEEEauthorblockA{
		\!\!\IEEEauthorrefmark{3}%
		Dept.\;of\;Medical\;Physics\;and\,Acoustics\;and\;Cluster\;of\;Excellence\;Hearing4all,\;%
		University\,of\;Oldenburg,\;%
		Oldenburg,\;Germany\\
	}
	\IEEEauthorblockA{
		\IEEEauthorrefmark{4}%
		Department of Electronic Systems, Aalborg University,
		Aalborg, Denmark
	}
}

\IEEEspecialpapernotice{(Invited Paper)\vspace*{-2px}}
	
\maketitle

\vspace*{-4mm}%
\begin{abstract}
	Determining the head orientation of a talker is not only beneficial for various speech signal processing applications, such as source localization or speech enhancement, but also facilitates intuitive voice control and interaction with smart environments or modern car assistants.
	Most approaches for head orientation estimation are based on visual cues. However, this requires camera systems which often are not available.
	We present an approach which purely uses audio signals captured with only a few distributed microphones around the talker.
	Specifically, we propose a novel method that directly incorporates measured or modeled speech radiation patterns to infer the talker's orientation during active speech periods based on a cosine similarity measure.
	Moreover, an automatic gain adjustment technique is proposed for uncalibrated, irregular microphone setups, such as ad-hoc sensor networks.
	In experiments with signals recorded in both anechoic and reverberant environments, the proposed method outperforms state-of-the-art approaches, using either measured or modeled speech radiation patterns.
\end{abstract}

\vspace*{5pt}
\begin{IEEEkeywords}
	acoustic head orientation estimation, speech directivity, speech radiation pattern, distributed microphones
\end{IEEEkeywords}

\vspace*{2pt}%
\section{Introduction}

\noindent
Measurements of the human speech radiation pattern reveal that the average speech directivity is highly frequency dependent:
while low frequencies are emitted almost omnidirectionally, high frequencies are strongly directive and reach attenuations of more than 20\,dB above 8\,kHz in the backward direction compared to the frontward direction~\cite{Chu2002, Bellows2019, Leishman2021} (see Figs.\,\ref{fig:sd} and~\ref{fig:mag_diff}).
Due to this strong directivity, the orientation of a talker has an impact on speech intelligibility~\cite{Liang2023} and may be crucial to consider in some speech processing applications.
For instance, incorporation of the head orientation can improve speech enhancement~\cite{Chakrabarty2016, AlMafrachi2018} or speech source localization~\cite{Brutti2007}.
Additionally, knowledge of the talker's orientation facilitates intuitive voice control in smart environments involving smart device detection~\cite{Mueller2016, Yang2021} or navigation~\cite{Sasou2009}.

Compared to the large variety of approaches for head orientation estimation using visual features~\cite{MurphyChutorian2009}, those using only audio features are scarce.
However, due to the omnipresence of smart devices around us, each equipped with at least one microphone, audio-based approaches involving multiple spatially distributed microphones are becoming more desirable and relevant in practice.
\pagebreak

\vspace*{-4mm}
Early works introduced the Oriented Global Coherence Field (OGCF) to jointly estimate the speaker's position and head orientation using a variation of the steered response power with phase transform (SRP-PHAT) algorithm~\cite{Brutti2005, Brutti2007} with distributed microphone arrays.
Several publications applying similar approaches or extensions thereof followed~\cite{Svaizer2012, Segura2014, Mueller2016}.
Other methods for head orientation estimation implicitly exploit the characteristic speech directivity in different frequency bands with distributed arrays~\cite{Abad2007, Segura2008, Felsheim2021} or a single array~\cite{Barnard2016}.
By contrast, \cite{Nakajima2009, Levi2010} use datasets of measured source radiation patterns rather than heuristic features to deduce the speaker orientation by matching an observed radiation pattern with the measured datasets using extensive microphone arrays.
Other publications apply machine learning techniques~\cite{Nakano2009, Takashima2011, Takashima2012, AlMafrachi2018}, often aiming at reducing the number of microphones involved.

In this work, we propose a novel method to estimate the talker's head orientation with only a few distributed single microphones, where the angles of the microphones relative to the talker are known a priori.
Specifically, we compute the cosine similarity between the signal power captured at the microphones and expected energy patterns for multiple candidate head orientations.
To this end, the expected energy patterns can be extracted from either measured or modeled speech radiation patterns.
Furthermore, we present a method to compensate for distance or gain deviations between microphones, which also makes it applicable to ad-hoc networks.
An evaluation study was performed involving recordings of different speaker positions and multiple orientations with six distributed microphones in an anechoic room and a car interior.
The proposed method is compared to applicable state-of-the-art methods for both cases: using measured or modeled speech radiation patterns.

\vspace*{3pt}%
\section{Directivity of Human Speech}
\label{sec:radiation_patterns}

\noindent
The radiation of human speech differs significantly between low frequencies, where sound is emitted almost omnidirectionally, and high frequencies with strong directivity towards the frontal direction~\cite{Leishman2021} (see Fig.\,\ref{fig:sd}).
There are several publications that comprise a detailed directional analysis of human or artificial speech radiation based on measurements~\cite{Chu2002, Halkosaari2005, Bellows2019, Bellows2019a, Leishman2021, Poerschmann2020}.
Other works focus on the modeling of the mouth radiation or compare such analytic models with measurements of human or artificial speech~\cite{Chalker1985, Huopaniemi1999, Fischer2019}.
Such models usually consist of a vibrating mouth opening in a rigid sphere (head), whereas the torso is not modeled.
Below, the datasets and the analytic speech radiation model which we used for the proposed head orientation estimation approach (see Section~\ref{sec:proposed_HOE}) are briefly described.

\subsection{Datasets of Measured Speech Directivities}

\noindent
\vspace*{-5mm}%
\begin{itemize}
	\item[\cite{Chu2002}\hspace{1px}:]
	Speech directivities averaged over 40 human talkers and directivity of a B\&K 4128 head and torso simulator (HATS) in 1/3-octave bands between 160\,Hz and 8\,kHz.
	\item[\cite{Bellows2019}\hspace{1px}:]
	Speech directivities averaged over 6 human talkers and directivity of a KEMAR HATS in 1/3-octave bands between 100\,Hz and 10\,kHz with finely sampled angular grid.
	Fig.\,\ref{fig:sd} shows the according speech radiation patterns in the horizontal and sagittal plane for different frequencies.
\end{itemize}

\vspace*{3pt}%
\subsection{Analytic Model of Speech Directivity}

\noindent
In~\cite{Fischer2019}, a physical model for speech radiation is described and the resulting directivity is compared to measured directivities of a HATS.
The mouth and head are represented by a circular piston in a spherical baffle, which yields a rotationally symmetric radiation pattern.
According to~\cite{Williams1999}, the sound pressure in the free field at a point with distance~$r$ from the source at an-\mbox{gle\;$\theta$\;with\;respect\;to\;the\;frontal direction can be calculated as}
\begin{align}
	p(\omega, r, \theta) 
	&=
	\frac{i \rho_0 c \, v_0}{2}
	\,
	\sum_{n=0}^{\infty}
	\kappa_n
	\frac{h_n^{(2)}\?\big(\frac{\omega}{c} r \big)}{
		{{h_n\!\!\!}'^{(2)}}\?\big(\frac{\omega}{c} a \big)} \,
	P_n(\cos \theta) \,,
	\label{eq:analytic_model}
\end{align}
where $\omega \hspace*{2pt}{=}\hspace*{2pt} 2\pi f$ is the angular frequency, 
$i$ is the imaginary unit,
$\rho_0$ is the fluid density of air,
$c$ is the speed of sound,
$v_0$ is the velocity of the piston,
$h^{(2)}_n\!$ in the numerator is the spherical Hankel function of second kind of order $n$\footnote{The wave number, mostly denoted by $k$, is written explicitly as~$\frac{\omega}{c}$ here to avoid confusion with the frequency bin index that is used later.}, while the denominator contains the derivative of $h^{(2)}_n\!$, $a$ is the radius of the spherical head model, 
$P_n$ is the Legendre polynomial of order $n$ and
$\kappa_n \shrt{=} [ P_{n-1}(\cos \alpha) - P_{n+1}(\cos \alpha) ]$,
where $\alpha$ is the angle defining the radius of the circular piston (mouth opening) and $\kappa_{0} \hspace*{2pt}{=}\hspace*{2pt} 1-\cos \alpha$.
The directivity at angle $\theta$ relative to the forward direction $(0^\circ)$ with $r \gg a$ can be calculated as
\begin{align}
	D(\omega, \theta) = \frac{p(\omega, r, \theta)}{p(\omega, r, 0^\circ)}.
\end{align}
In this work we used a head radius $a \shrt{=}$9\,cm, and $\alpha \shrt{=} $5.7$^\circ$ corresponding to a mouth opening with 9\,mm radius.
The infinite sum in Eq.\,\eqref{eq:analytic_model} was evaluated until order $n\lt{max} \shrt{=} $50.

\vspace*{4pt}%
\subsection{Comparison of Measured and Modeled Directivities}
\label{sec:comparison_model_measured}

\begin{figure}[tbp]
	\centering
	\includegraphics[width=0.995\linewidth]{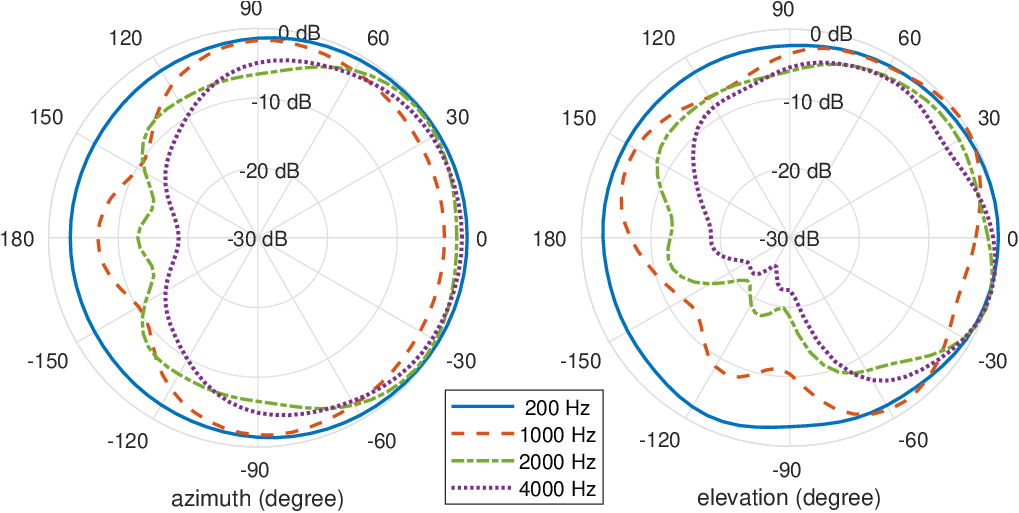}
	\\
	\hspace*{0pt}
	\begin{subfigure}[b]{0.45\linewidth}
		\centering
		\caption{Horizontal plane (elevation 0°)}
		\label{fig:sd_horizontal}
	\end{subfigure}%
	\hfill
	\begin{subfigure}[b]{0.45\linewidth}
		\centering
		\caption{Vertical, sagittal plane (azimuth 0°)}
		\label{fig:sd_sagittal}
	\end{subfigure}
	\caption{.~~Average radiation patterns of human speech for different frequencies according to~\cite{Bellows2019}. The right-hand side (0$^\circ$) indicates the front direction.}
	\label{fig:sd}
	\vspace*{3mm}%
	\includegraphics[width=0.995\linewidth, trim=0mm 0mm 1mm 0mm, clip]{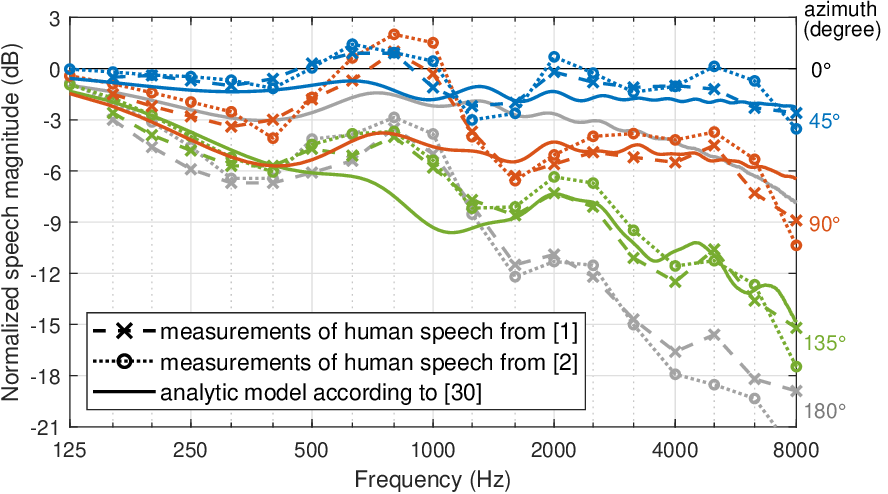}%
	\vspace*{-1pt}%
	\caption{.~~Speech magnitudes in 1/3-octave bands for different azimuth angles between 0$^\circ$ and 180$^\circ$ in the horizontal plane normalized to the front direction (0$^\circ$). Dashed and dotted lines refer to average measured data from~\cite{Chu2002, Bellows2019}. Solid lines display magnitudes according to an analytic radiation model~\cite{Williams1999}.}
	\label{fig:mag_diff}
	\vspace*{-1mm}%
\end{figure}

\noindent
Fig.\,\ref{fig:mag_diff} compares the horizontal frequency-dependent speech directivity of the reference measurements and the analytic model for different azimuth angles.
While the measured directivity data of~\cite{Chu2002} and~\cite{Bellows2019} are similar, the analytic model shows some distinct deviations from the measured data due to the missing torso, especially between 500\,Hz and 1500\,Hz.
Furthermore, the model has much less attenuation towards the rear direction (180$^\circ$), where only less than 8\,dB appear at 8\,kHz, compared to about 20\,dB for the measured data.

\section{Head Orientation Estimation Using\linebreak Individual Microphone Features}
\label{sec:hoe_features}

\noindent
In this section, we review applicable state-of-the-art methods for head orientation estimation.
Specifically, we only consider reference methods that can be applied using distributed microphones (i.e.\ not distributed arrays) and require no training.
The approaches described below estimate the head orientation in two steps:
head orientation information is firstly obtained by individual microphone features and, secondly, these features are used in a vectorial orientation decision technique.

\vspace*{2pt}%
\subsection{Individual Microphone Features}
\label{sec:features}

\subsubsection*{High-to-Low-Band Energy Ratio (HLBR)\normalfont{\cite{Abad2007, Segura2008}}}

This method introduces a feature that computes the energy ratio between a high and low frequency band.
Two different head orientation estimation methods based on this feature are described in~\cite{Abad2007}. 
We consider the vectorial HLBR here, which showed better performance at lower complexity.
For each microphone $m \shrt{\in} \{0, ..., M{-}1\}$, the HLBR is computed as
\begin{align}
	\operatorname{HLBR}_m
	&=
	\frac{\sum_{k \in \mathcal{K}_\text{hi}} | Y_m(k) |^2 }{
		\sum_{k \in \mathcal{K}\lt{lo}}  | Y_m(k) |^2 } \,,
\end{align}
where $Y_m(k)$ is the short-time Fourier transform (STFT) spectrum of microphone~$m$ at frequency bin~$k$.
The frame index is omitted for better readability.
The sets $\mathcal{K}\lt{hi}$ and $\mathcal{K}\lt{lo}$ contain all frequency bins of the high and low frequency band, respectively.
Tab.\;\ref{tab:freq_bands} explicitly displays the frequency bands used in this work for each of the described approaches\footnote{The high- and low-band frequencies differ between~\cite{Abad2007, Segura2008, Felsheim2021}. We chose the frequencies yielding the best results in the present evaluation study.}.
As high frequencies are highly attenuated at a microphone behind the talker, the resulting HLBR can be expected to be lower than that of a microphone in front of the talker.
However, speech is highly dynamic and often there is primarily speech energy in either the low or the high frequency band.
In practice, this may lead to strong fluctuations of the HLBR measure.
The HLBR is independent of individual microphone gains and thus requires no microphone calibration.
\\[-4pt]

\subsubsection*{High-Band Variance (HBV)\normalfont{\cite{Felsheim2021}}}

Similar to the previous feature, the HBV makes use of the increasing low-pass characteristics of speech radiation towards rear directions.
Accordingly, the variance of high-frequency speech spectra behind the talker is assumed to be lower than in front of the talker.
Though originally proposed for distributed microphone arrays, this feature is directly applicable to individual microphones as
\begin{align}
	\operatorname{HBV}_m
	&=
	\frac{1}{d_m} \,
	\bigg(
	\frac{1}{|\mathcal{K}\lt{hi}|} \;\,
	\sum_{\mathclap{k \in \mathcal{K}\lt{hi}}} \big( 
	| Y_m(k) | - \mu_m
	\big)^2 \bigg) ,
\end{align}
{where\;$\mu_m \hspace*{1px}{=}\hspace*{1px} \frac{1}{|\mathcal{K}\lt{hi}|} \!\sum_{k \in \mathcal{K}\lt{hi}} \!|Y_m(k)|$\;is\;the mean magnitude spec-}trum over a high frequency band $\mathcal{K}\lt{hi}$, and $|\mathcal{K}\lt{hi}|$ denotes the number of frequency bins in this band. The same notation ($|\cdot|$) will be used to refer to the cardinality of other sets in the remainder of this paper.
The HBV is weighted by the inverse of the distance $d_m$ between the talker and microphone~$m$ in order to de-emphasize the variance of microphones with larger distance to the talker, where the sound field is more diffuse.
This feature thus requires knowledge of the microphone distances.
Furthermore, it is sensitive to different microphone gains as it is proportional to the microphone signal power.\\[-4pt]

\subsubsection*{Spectral Difference (SD)\normalfont{\cite{Felsheim2021}}}

In contrast to the previous features, which exploit information of each microphone individually, the SD relates individual high-band magnitude spectra to an average of all microphones.
Furthermore, the authors propose a cepstral smoothing of the microphone spectra to reduce the influence of noise and the speech content.
The SD is defined as
\begin{align}
	\operatorname{SD}_m
	&=
	\frac{1}{|\mathcal{K}\lt{hi}|} \;\;
	\sum_{\mathclap{k \in \mathcal{K}\lt{hi}}}
	\big( |\check{Y}_m(k)| - {\check{Y}}\lt{mean}(k) \big)\,,
\end{align}
where $\check{Y}_m(k)$ denotes the spectrum after cepstral low-pass
\mbox{liftering and\;$\check{Y}\lt{mean}(k) \hspace*{1pt}{=}\hspace*{1pt} \frac{1}{M} \sum_{m} \!|\check{Y}_m(k)|$\;is the average magni-} tude spectrum of all $M$ microphones.
Similar to the HBV, this feature is influenced by individual microphone gains and thus necessitates microphone calibration.

\vspace*{6pt}
\subsection{Vectorial Head Orientation Decision}
\label{sec:vectHOdecision}

\noindent
This geometrical technique for head orientation estimation based on individual microphone features was initially proposed in~\cite{Abad2007}.
The head orientation angle $\hat{\theta}$ can be estimated by sum- ming over feature-weighted microphone steering vectors as
\pagebreak
\vspace*{-5mm}%
\begin{align}
	\hat{\theta}
	&=
	\angle \m{v}\lt{sum} \;\;
	\text{with } \;\;
	\m{v}\lt{sum} = \sum_{\mathclap{m=0}}^{\mathclap{M-1}}
	\m{v}_m \, \{ \bullet \}_m\,,
\end{align}
where $\m{v}_m$ is a unit-length vector pointing from the talker to microphone $m$, $\{ \bullet \}_m$ is a placeholder for the according feature ($\operatorname{HLBR}_m$, $\operatorname{HBV}_m$ or $\operatorname{SD}_m$), and $\angle$ denotes the azimuth with respect to a reference vector defining the frontal orientation~0°.
To reduce fluctuations of $\hat\theta$ caused by the non-stationarity of speech, recursive smoothing can be applied to the microphone STFT spectra $Y_m(k)$.

A practical drawback of the vectorial head orientation decision method is that it implicitly assumes a uniform angular distribution of the microphones surrounding the talker.
In practice, this approach thus might be sensitive to off-center talker positions or setups with uneven angular microphone distribution, as becoming evident in the evaluations in Section~\ref{sec:evaluations}.

\vspace*{8pt}%
\section{Proposed Head Orientation Estimation\linebreak Using Speech Radiation Patterns}
\label{sec:proposed_HOE}

\noindent
In this section, the proposed method for head orientation estimation is described.
Unlike the features described in the previous section, which heuristically exploit the directivity of high-frequency speech in a single frequency band, this approach explicitly uses speech radiation patterns in multiple frequency bands to estimate the talker's orientation.

\vspace*{3pt}%
\subsection{Radiation Pattern Matching (RAPM)}
\label{sec:RAPM}

\noindent
We propose to evaluate the short-time power spectral density (PSD) of each microphone $m$ in 1/3-octave bands:\
\begin{align}
	\varPhi_{m,b}
	&=
	\frac{1}{|\mathcal{K}_b|} \;
	\sum_{\mathclap{k \in \mathcal{K}_b}} \varPhi_{m}(k) \,,
	\label{eq:mic_PSD_b}
	\vspace*{-2pt}
\end{align}
where $b$ is the frequency band index, $\varPhi_{m}(k) = \Ex\big\{ |Y_m(k)|^2 \big\}$ is the short-time PSD with $\Ex \{ \cdot \}$ being the expectation operator, and $\mathcal{K}_b$ is a set containing all frequency bins of 1/3-octave band $b$.
In practice, PSDs can be estimated by recursive smoothing of instantaneous PSDs $|Y_m(k)|^2$ in each time frame.
From the speech radiation patterns described in Section~\ref{sec:radiation_patterns}, we can directly extract the expected speech radiation power towards each microphone direction for any hypothetical head orientation $\theta$ as
\begin{align}
	P_{m,b}(\theta)
	&= \big|
	D_b(\theta_m\shrt{-}\theta, \varphi_m) \big|^2\,,
	\label{eq:P_m}
\end{align}
where $D_b(\theta_m\shrt{-}\theta, \varphi_m)$ is the speech directivity towards microphone $m$ with azimuth $\theta_m$ and elevation $\varphi_m$ in the $b$-th band, relative to a candidate head orientation $\theta$.\footnote{Note that even though the scope of this work is limited to horizontal head orientations, Eq.\,\eqref{eq:P_m} could be extended to consider head elevation as well.}
It should be noted that the orientation-dependent radiation power $P_{m,b}(\theta)$ is subject to certain assumptions, namely that (a) each microphone has the same distance to the talker, (b) each microphone has the same gain and sensitivity, and that (c) the microphones are in free field.
For now, it is assumed that these conditions are fulfilled. The following section addresses microphone setups violating these assumptions.

To estimate the head orientation, the observed microphone PSDs in Eq.\,\eqref{eq:mic_PSD_b} are compared to the expected radiation powers in Eq.\,\eqref{eq:P_m} for multiple head orientation candidates.
To this end, both quantities are initially stacked into vectors comprising all microphones:
\begin{align}
	{\m{\Phi}}_b
	&=
	\big[
	\varPhi_{0,b}, \, \dots, \, \varPhi_{M\?\?-\?1,b}
	\big]^\mathrm{T} ,
	\label{eq:vec_phi}
	\\
	\m{P}_{\?\?b}(\theta)
	&=
	{\big[ P_{0,b}(\theta), \dots, P_{\?\?M-\?1,b}(\theta) \big]}^\mathrm{T} .
	\label{eq:vec_P}
\end{align}
As cost function, we propose to evaluate the cosine similarity between the two vectors in each 1/3-octave band $b$, and average over a set of bands $\mathcal{B\lt{s}}$ that are relevant for speech, as
\begin{align}
	J({\theta})
	\?&= \?
	\frac{1}{|\mathcal{B}\lt{s}|}
	\sum_{b \in \mathcal{B}\lt{s}}\!
	\cos \angle \big(\m{\Phi}_b, \m{P}_{\?\?b}(\theta) \big)
	\?=\?
	\frac{1}{|\mathcal{B}\lt{s}|}
	\sum_{b \in \mathcal{B}\lt{s}}\!
	\frac{{\m{\Phi}}_b^\mathrm{T}}{\Vert {\m{\Phi}}_b \Vert}
	\frac{\m{P}_{\?\?b}(\theta)}{\Vert \vphantom{\tilde{\m{\Phi}}} \m{P}_{\?\?b}(\theta) \Vert},
	\label{eq:cost_function}
\end{align}
where $\cos \angle (\m{a}, \m{b})$ denotes the cosine of the intermediate angle of vectors $\m{a}$ and $\m{b}$.
Maximizing Eq.\,\eqref{eq:cost_function} finally yields the head orientation estimate
\begin{align}
	\hat{\theta}
	&=
	\arg \max_{\theta} \: J({\theta}) \,.
\end{align}
The cost function in Eq.\,\eqref{eq:cost_function} takes real values within the interval $[0,1]$, where a perfect match of the observed and expected power spectra results in $J\lt{max} \shrt{=} 1$.
Hence, the value of the cost function at the estimated head orientation $J(\hat\theta)$ additionally contains information about the quality of the match, which can be interpreted as an indication of confidence.

\vspace*{6pt}
\subsection{Microphone Gain Adjustment}
\label{sec:gain_adjustment}

\noindent
As mentioned before, the proposed radiation pattern matching is sensitive to systematic level differences between the microphones.
These might appear if the distance to the talker differs from one microphone to the other or if the microphones have different sensitivity or gain and are not calibrated.
In such situations, there is a systematic mismatch between the observed microphone PSDs from Eq.\,\eqref{eq:mic_PSD_b} and the expected radiation powers from Eq.\,\eqref{eq:P_m}.
To compensate for this, we propose a microphone gain adjustment pre-processing:
\begin{align}
	\tilde{\varPhi}_{m,b}
	&=
	a_m
	\,
	\varPhi_{m,b} \,,
	\label{eq:PSD_comp}
\end{align}
which scales the microphone PSDs with a real-valued gain~$a_m$ to correspond to a situation where the microphones would be calibrated and equidistant to the talker.
Here, we assume that relevant information about the head orientation is present in each microphone.
Otherwise, one could apply a microphone pre-selection, which is beyond the scope of the present work.
\\[-5pt]

\subsubsection*{Distance-dependent gain adjustment}
The most intuitive solution to compensate for different distances $d_m$ from the microphones to the talker might be to equalize the distance-dependent attenuation of the speech signal power with respect to a reference microphone, e.g., $m\shrt{=}$0, defining
\begin{align}
	a_{m,\text{dist}}
	&=
	\left( \frac{d_m}{d_0} \right)^2.
	\label{eq:ddga}
\end{align}
However, this gain adjustment is only valid in the free field, where the signal power decreases according to $1/d^2$.
In a reverberant environment, Eq.\,\eqref{eq:ddga} would lead to inaccurate compensation and overamplification of distant microphones.
Moreover, strong noise amplification in noisy, distant microphones appears to become problematic in practice as well.
Hence, we propose a different gain adjustment method below that considers the above-stated issues.\\[-4pt]

\subsubsection*{Low-frequency gain adjustment (LFA)}
This method exploits the knowledge that speech radiation is near-omnidirec- tional in low frequencies (see Fig.\,\ref{fig:sd}).
We would thus observe almost identical speech power in a low frequency band -- independent of the talker orientation -- in each microphone of an ideal microphone setup (i.e., with equal distance from the talker to calibrated  microphones in free field).
This approach attempts to restore the described ideal conditions by adjusting the microphone gains so that the low-frequency signal power in each microphone equals that of a reference microphone (e.g., $m\shrt{=}0$):
\vspace*{-4pt}%
\begin{align}
	a_{m,\text{LFA}}
	&=
	\frac{\sum_{b \in \mathcal{B}\lt{LFA}} \bar{\varPhi}_{0,b}}{
		\sum_{b \in \mathcal{B}\lt{LFA}} \bar{\varPhi}_{m,b}} \,.
	\label{eq:LFA}
\end{align}
Here, $\mathcal{B}\lt{LFA}$ is a set of low-frequency 1/3-octave bands.
To minimize speech-signal-related fluctuations of $a_{m,\text{LFA}}$, we use long-time averaged PSDs $\bar{\varPhi}_{m,b}$, which are updated during frames with active speech (cf.\ Tab.\;\ref{tab:freq_bands} and \ref{tab:algorithmic_params} for details).

Besides compensating for microphone distance deviations also in reverberant environments, the proposed gain adjustment implicitly calibrates microphones with different sensitivities (assuming identical frequency responses), which makes it specifically applicable to ad-hoc microphone networks.
Moreover, it is sufficient to know the microphone directions relative to the talker position, whereas the distance information of the microphones is not required.

\begin{table}[h]
	\centering
	\caption{\centering \linebreak \textsc{Frequency bands for the presented methods.}}
	\vspace*{-1.5mm}%
	\begin{tabular}{lccl}
		\hline
		\multicolumn{1}{c}{HLBR}         & \rule[-0.4\normalbaselineskip]{0pt}{1.4\normalbaselineskip} HBV                        & SD                       & \multicolumn{1}{c}{RAPM}       \\
		\hline
		\rule{0pt}{1.1\normalbaselineskip}%
		\!$\mathcal{K}\lt{lo}$: 200\shrt{-}400\,Hz & \multirow{2}{*}{\hspace*{-2px}$\mathcal{K}\lt{hi}$: 5\shrt{-}8\,kHz} & \multirow{2}{*}{\hspace*{-2px}$\mathcal{K}\lt{hi}$: 5\shrt{-}8\,kHz} & \;\;\,$\mathcal{B}\lt{s}$: \; 1\shrt{-}8\,kHz \\
		\!$\mathcal{K}\lt{hi}$: \;\;4\shrt{-}8\,kHz    &                            &                          & \hspace*{-2px}$\mathcal{B}\lt{LFA}$:\ 100\shrt{-}400\,Hz\hspace*{-5px}
		\rule[-0.4\normalbaselineskip]{0pt}{0pt} \\
		\hline
	\end{tabular}
	\label{tab:freq_bands}
	\vspace*{1mm}
\end{table}

Note that a gain adjustment is not only applicable to the proposed radiation pattern matching: the same technique can be applied to scale the microphone spectra $Y_m(k)$ before computing the individual microphone features described in Section\;\ref{sec:features}.
Specifically, the HBV and SD, which are sensitive to systematic microphone level differences and require cali- bration, might benefit from a gain adjustment pre-processing.
In order to demonstrate the effect of the proposed low-frequency gain adjustment (LFA) from Eq.\,\eqref{eq:LFA}, the evaluations in the following section comprise results for the presented head orientation estimation techniques both without gain adjustment and with LFA.

\vspace*{6pt}%
\section{Experiments and Results}
\label{sec:evaluations}

\begin{figure}[h]
	\centering
	\hspace*{4mm}%
	\begin{subfigure}[b]{0.85\linewidth}
		\includegraphics[width=1\linewidth, trim=0mm 104mm 0mm 0mm, clip]{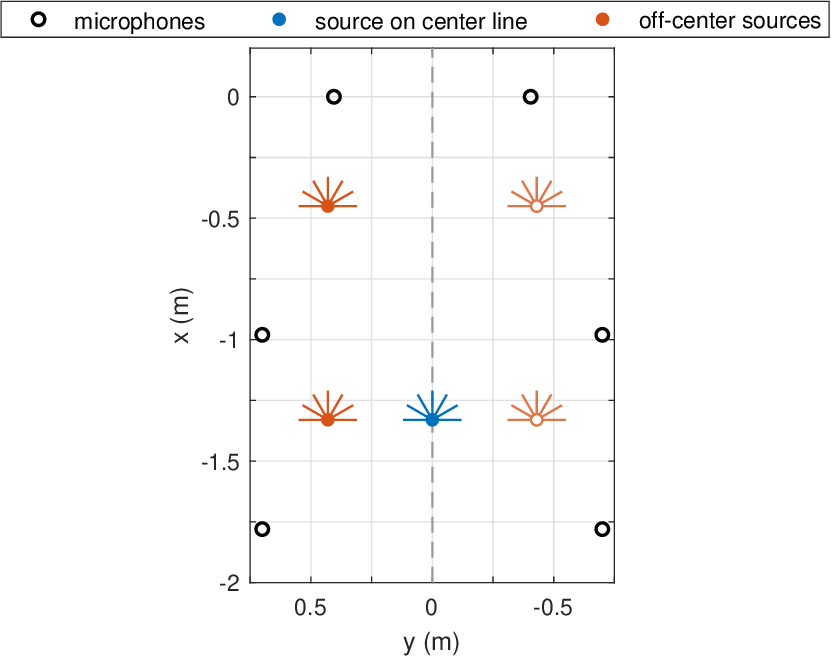}
	\end{subfigure}
	\\[2pt]
	\begin{subfigure}[b]{0.49\linewidth}
		\hspace*{-2mm}%
		\includegraphics[width=0.89\linewidth, trim=0mm 0mm 0mm 0mm, clip]{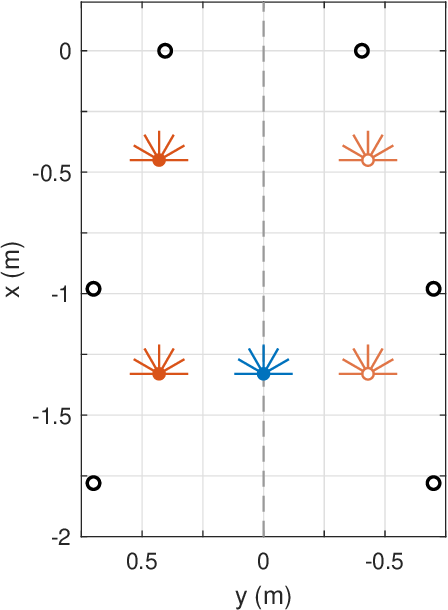}
		\centering
		\vspace*{-2pt}%
		\caption{Setup in anechoic room (experiment\,A).}
		\label{fig:setup_AAU}
	\end{subfigure}%
	\hfill
	\begin{subfigure}[b]{0.49\linewidth}
		\hspace*{-2mm}%
		\includegraphics[width=0.89\linewidth, trim=0mm 0mm 0mm 0mm, clip]{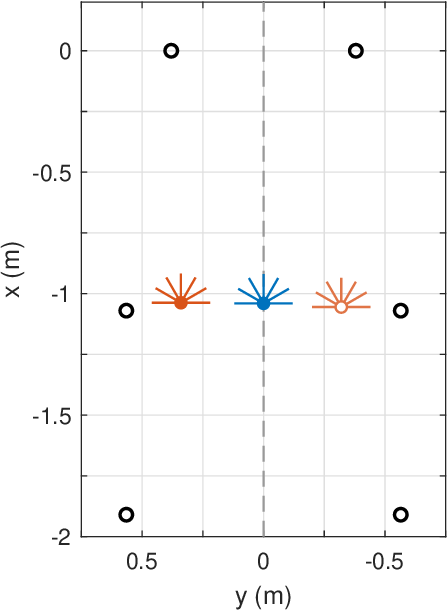}
		\centering
		\vspace*{-2pt}%
		\caption{Setup in car interior (experiment\;B).}
		\label{fig:setup_V}
	\end{subfigure}%
	\vspace*{-2pt}%
	\caption{.~~Geometry of the experiments. Microphones are elevated 25\,cm above the mouth. Lines around the talker positions show evaluated head orientations.}
	\label{fig:setup}
\end{figure}

\noindent
We performed an evaluation study in two different environments involving six distributed microphones in a similar setup.
Fig.\,\ref{fig:setup} shows the microphone and talker positions as well as the evaluated head orientation angles of both experiments.

\begin{table}[h]
	\centering
	\caption{\centering \linebreak \textsc{stft and smoothing parameters used for evaluations.}}
	\vspace*{-1.6mm}%
	\begin{tabular}{cccccc}
		\hline
		\rule{0pt}{1.1\normalbaselineskip}%
		sample rate &  frame size  & hop size    & window & $\tau\lt{PSD}$ & $\tau\lt{LFA}$ \\
		16\,kHz    & 512\,samples & 256\,samples & Hann   & 250\,ms & 5\,s%
		\rule[-0.4\normalbaselineskip]{0pt}{0pt} \\ \hline
	\end{tabular}
	\label{tab:algorithmic_params}
	\vspace*{-3mm}
\end{table}

\subsubsection*{Experiment A}
The first experiment was conducted with measurements obtained in an anechoic room.
The microphone and source positions were set up to imitate the geometry of a van with microphones in the imaginary vehicle ceiling involving five different talker positions.
At each position, impulse responses were measured for head orientations between -90$^\circ$\,and 90$^\circ$\,in steps of\;30$^\circ$\,with the artificial mouth of a B\&K 4128 HATS.
The speech signals for each position and orientation were rendered by convolution of the measured impulse responses with 3\,s clean speech snippets from a database~\cite{Graetzer2022} (5 female, 5 male subjects).
Pink noise signals were generated with a simulation toolbox for isotropic noise fields~\cite{Habets2007} at 10\,dB SNR (averaged over all microphones).
\\[-8pt]

\subsubsection*{Experiment B}
The second experiment involved recordings from a car interior (Mercedes V-class) with reverberation time $T_{60}\shrt{=}$90\,ms.
Six seconds long speech segments spoken by a human, male talker were recorded at three positions\;and seven head orientations.
Driving noise at constant speed was recorded separately resulting in an average SNR of 12.4\,dB.
\\[-8pt]

\begin{figure}[htp]
	\centering
	\begin{subfigure}[b]{1\linewidth}
		\centering
		\includegraphics[width=1\linewidth]{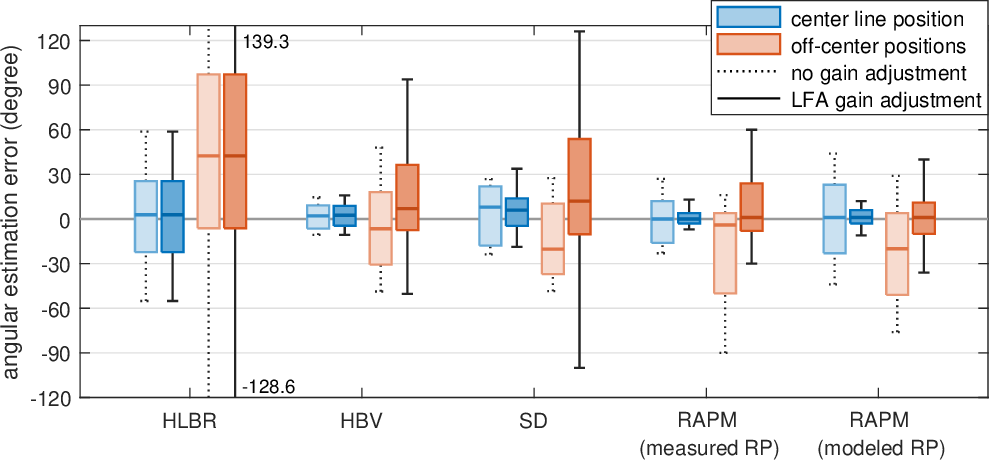}
		\vspace*{-5mm}%
		\caption{Results in the anechoic room (experiment A) with pink noise at 10\,dB SNR.}
		\label{fig:results_AAU}
	\end{subfigure}
	\\
	\vspace*{3mm}%
	\begin{subfigure}[b]{1\linewidth}
		\centering
		\includegraphics[width=1\linewidth]{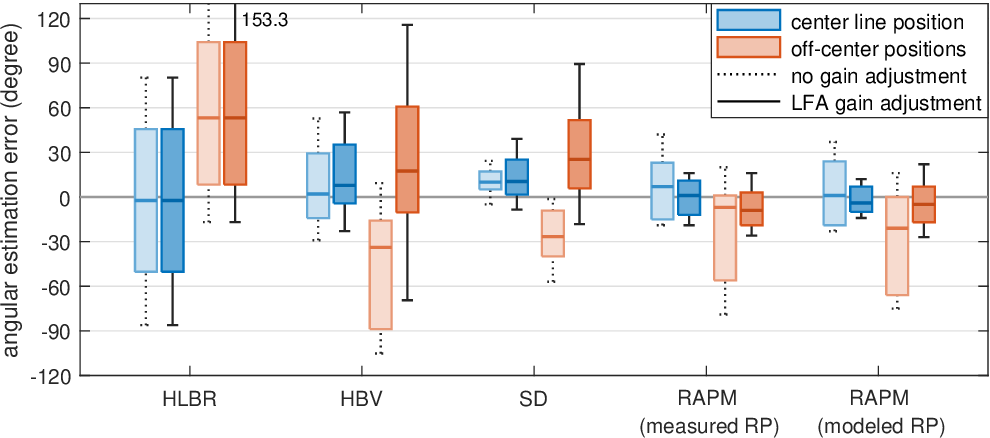}
		\vspace*{-5mm}%
		\caption{Results in the car interior (experiment B) with driving noise at 12.4\,dB SNR.}
		\label{fig:results_V}
	\end{subfigure}%
	\caption{.~~%
		Distribution of the angular head orientation estimation error for state-of-the-art methods (HLBR, HBV, SD) and the proposed radiation pattern matching (RAPM) involving measured or modeled speech radiation patterns (RP).
		Light box plots with dashed whisker lines correspond to results without low-frequency gain adjustment (LFA); dark box plots with solid whisker lines correspond to results with LFA.
		Boxes indicate the median and lower\,/\,upper quartile error value.
		The whiskers indicate the 10\%\,/\,90\% quantile, where whiskers outside the plotted area are quantified by numbers.}
	\label{fig:results}
	\vspace*{-3mm}
\end{figure}

\subsubsection*{Angular estimation error}
At all time frames with active speech, we computed the angular head orientation estimation error of the reference methods (HLBR, HBV, SD) described in Section\;\ref{sec:features}  using the vectorial orientation decision (cf.\ Section~\ref{sec:vectHOdecision}), and the proposed radiation pattern matching (RAPM, cf.\ Section\;\ref{sec:RAPM}) using either measured\footnote{Measured directivities from B\&K HATS~\cite{Chu2002} were used for experiment~A, and average directivities from human test subjects~\cite{Bellows2019} for experiment B.} or modeled speech directivities.
Additionally, all methods were evaluated without and with low-frequency gain adjustment (LFA, cf.\ Section\;\ref{sec:gain_adjustment}).
The used algorithmic parameters for both experiments are listed in Tab.\;\ref{tab:algorithmic_params}, where $\tau\lt{PSD}$ and $\tau\lt{LFA}$ are the time constants used for recursive smoothing in order to estimate the PSD in Eq.\,\eqref{eq:mic_PSD_b} and the long-term PSD in Eq.\,\eqref{eq:LFA}, respectively.
Furthermore, the right-hand side source positions (indicated by hollow circles in Fig.\,\ref{fig:setup}) were mirrored on the center line to the left-hand side, and the angular errors of all off-center positions were pooled in order to indicate systematic estimation errors of unilateral off-center positions.
\\[-8pt]

\subsubsection*{Results and discussion}
Fig.\,\ref{fig:results} shows the distribution of the angular estimation error of both experiments over all time frames with active speech and all orientations. 
The results for the center-line position and all off-center source positions are separately displayed.
While the proposed LFA pre-processing shows no clear improvement of the reference methods HBV and SD (no influence on scaling-invariant HLBR), the performance of the proposed RAPM clearly benefits from the LFA and outperforms the reference methods in all algorithmic configurations.
At the center line positions, the angular estimation errors mostly are evenly distributed around zero with a median close to zero degree, whereby RAPM consistently shows the lowest error values.
At off-center positions, a systematic bias of the estimated orientation is observable for HLBR, HBV and SD without LFA, especially in the car interior (experiment B).
With LFA, the estimation bias can be reduced slightly, however, the error variance increases.
By contrast, the proposed RAPM (with LFA) has no significant bias at off-center positions and shows the smallest error range.
No significant differences in performance are observable between using measured or modeled speech radiation patterns in both experiments.
The reason for this might be that the analytic speech radiation model matches the horizontal human speech radiation pattern well for most frequencies except for frequencies between 500 and 1500\,Hz and a narrow side lobe at the rear direction (cf.\ Section\;\ref{sec:comparison_model_measured}).
However, relevant differences between the model and measurements might appear for other setups, especially when the shadowing effect of the torso has more impact, e.g., when microphones are below the head.

\vspace*{5pt}%
\section{Conclusion}

\noindent
We presented a method to estimate the head orientation of a human talker using a few distributed microphones around the talker.
Unlike comparable state-of-the-art approaches, which make use of the speech directivity in a high frequency band based on individual microphone features, the proposed method directly exploits speech radiation patterns by extracting the expected radiation power towards each microphone for multiple head orientation candidates.
The talker's orientation is then assessed by matching the orientation-dependent expected microphone power with the observed microphone power in multiple frequency bands (radiation pattern matching).
Moreover, we proposed an automatic gain adjustment to compensate for different microphone sensitivities or distances from the talker, specifically in reverberant environments.
This method implicitly eliminates the need for microphone calibration and requires no information about the microphone distances.

We evaluated the proposed method in experiments involving anechoic and reverberant speech signals.
In particular, we compared the performance to comparable state-of-the-art methods using measured or modeled speech radiation patterns.
At all evaluated positions, but especially at off-center positions, the proposed method outperforms the reference approaches.
However, no clear preference of using speech radiation patterns from measurements or those extracted from a simple speech radiation model are observable in the evaluated setups.

Future work could comprise evaluations with fewer microphones and asymmetrical constellations, and an automatic microphone selection to disregard noisy or distant microphones.

\vspace*{5pt}%
\input{asilomar23_bbl.bbl}

\end{document}

%% file: asilomar23_bbl.bbl

%% file: asilomar23_paper.bbl
\begin{thebibliography}{10}
\providecommand{\url}[1]{#1}
\csname url@samestyle\endcsname
\providecommand{\newblock}{\relax}
\providecommand{\bibinfo}[2]{#2}
\providecommand{\BIBentrySTDinterwordspacing}{\spaceskip=0pt\relax}
\providecommand{\BIBentryALTinterwordstretchfactor}{4}
\providecommand{\BIBentryALTinterwordspacing}{\spaceskip=\fontdimen2\font plus
\BIBentryALTinterwordstretchfactor\fontdimen3\font minus
  \fontdimen4\font\relax}
\providecommand{\BIBforeignlanguage}[2]{{%
\expandafter\ifx\csname l@#1\endcsname\relax
\typeout{** WARNING: IEEEtran.bst: No hyphenation pattern has been}%
\typeout{** loaded for the language `#1'. Using the pattern for}%
\typeout{** the default language instead.}%
\else
\language=\csname l@#1\endcsname
\fi
#2}}
\providecommand{\BIBdecl}{\relax}
\BIBdecl

\bibitem{Chu2002}
W.~T. Chu and A.~C.~C. Warnock, ``Detailed directivity of sound fields around
  human talkers,'' Tech. Rep., 2002.

\bibitem{Bellows2019}
S.~D. Bellows, C.~M. Pincock, J.~K. Whiting, and T.~W. Leishman, ``Average
  speech directivity,'' Tech. Rep.~1, 2019.

\bibitem{Leishman2021}
T.~W. Leishman, S.~D. Bellows, C.~M. Pincock, and J.~K. Whiting,
  ``High-resolution spherical directivity of live speech from a
  multiple-capture transfer function method,'' \emph{J. Acoust. Soc. Am.}, vol.
  149, no.~3, pp. 1507--1523, Mar. 2021.

\bibitem{Liang2023}
L.~Liang and G.~Yu, ``Effect of speaker orientation on speech intelligibility
  in an automotive environment,'' \emph{Appl. Acoust.}, vol. 205, p. 109269,
  Mar. 2023.

\bibitem{Chakrabarty2016}
S.~Chakrabarty, D.~Pilakeezhu, and E.~A.~P. Habets, ``Head-orientation
  compensation with video-informed single channel speech enhancement,'' in
  \emph{2016 {IEEE} Int. Workshop Acoust. Signal Enhanc. ({IWAENC})}, Sep.
  2016.

\bibitem{AlMafrachi2018}
R.~Al-Mafrachi, M.~Gimm, and G.~Schmidt, ``Acoustic estimation of the head
  orientation for in-car communication systems,'' in \emph{Fortschritte der
  Akustik -- DAGA 2018}, Mar. 2018, pp. 1780--1783.

\bibitem{Brutti2007}
A.~Brutti, M.~Omologo, P.~Svaizer, and C.~Zieger, ``Classification of acoustic
  maps to determine speaker position and orientation from a distributed
  microphone network,'' in \emph{2007 {IEEE} Int. Conf. Acoustics, Speech and
  Signal Process ({ICASSP})}, Apr. 2007.

\bibitem{Mueller2016}
M.~M\"uller, S.~v.~d. Par, and J.~Bitzer, ``Head-orientation-based device
  selection: Are you talking to me?'' in \emph{Speech Communication; 12. ITG
  Symp.}, 2016, pp. 1--5.

\bibitem{Yang2021}
Q.~Yang and Y.~Zheng, ``Model-based head orientation estimation for smart
  devices,'' \emph{Proc. {ACM} Interactive, Mobile, Wearable and Ubiquitous
  Technologies}, vol.~5, no.~3, pp. 1--24, Sep. 2021.

\bibitem{Sasou2009}
A.~Sasou, ``Acoustic head orientation estimation applied to powered wheelchair
  control,'' in \emph{Proc. 2\textsuperscript{nd} Int. Conf. Robotic
  Communication and Coordination}, 2009.

\bibitem{MurphyChutorian2009}
E.~Murphy-Chutorian and M.~Trivedi, ``Head pose estimation in computer vision:
  A survey,'' \emph{{IEEE} Trans. Pattern Anal. Mach. Intell.}, vol.~31, no.~4,
  pp. 607--626, Apr. 2009.

\bibitem{Brutti2005}
A.~Brutti, M.~Omologo, and P.~Svaizer, ``Oriented global coherence field for
  the estimation of the head orientation in smart rooms equipped with
  distributed microphone arrays,'' in \emph{{Interspeech} 2005 -- Eurospeech,
  9\textsuperscript{th} Eur. Conf. Speech Commun. and Technology}, Sep. 2005, pp. 2337--2340.

\bibitem{Svaizer2012}
P.~Svaizer, A.~Brutti, and M.~Omologo, ``Environment aware estimation of the
  orientation of acoustic sources using a line array,'' in \emph{2012 Proc.
  20\textsuperscript{th} Eur. Signal Process. Conf. (EUSIPCO)}, 2012, pp.
  1024--1028.

\bibitem{Segura2014}
C.~Segura and J.~Hernando, ``3d joint speaker position and orientation tracking
  with particle filters,'' \emph{Sensors (Basel)}, vol.~14, no.~2, pp.
  2259--2279, Jan. 2014.

\bibitem{Abad2007}
A.~Abad, C.~Segura, C.~Nadeu, and J.~Hernando, ``Audio-based approaches to head
  orientation estimation in a smart-room,'' in \emph{Interspeech 2007}, Aug.
  2007.

\bibitem{Segura2008}
C.~Segura, A.~Abad, J.~Hernando, and C.~Nadeu, ``Speaker orientation estimation
  based on hybridation of {GCC-PHAT} and {HLBR},'' in \emph{Interspeech 2008},
  Sep. 2008.

\bibitem{Felsheim2021}
R.~C. Felsheim, A.~Brendel, P.~A. Naylor, and W.~Kellermann, ``Head orientation
  estimation from multiple microphone arrays,'' in \emph{2020
  28\textsuperscript{th} Eur. Signal Process. Conf. ({EUSIPCO})}, Jan. 2021,
  pp. 491--495.

\bibitem{Barnard2016}
M.~Barnard and W.~Wang, ``Audio head pose estimation using the direct to
  reverberant speech ratio,'' \emph{Speech Commun.}, vol.~85, pp. 98--108, Dec.
  2016.

\bibitem{Nakajima2009}
H.~Nakajima, K.~Kikuchi, T.~Daigo, Y.~Kaneda, K.~Nakadai, and Y.~Hasegawa,
  ``Real-time sound source orientation estimation using a 96 channel microphone
  array,'' in \emph{2009 {IEEE}/{RSJ} Int. Conf. Intelligent Robots and
  Systems}, Oct. 2009.

\bibitem{Levi2010}
A.~Levi and H.~Silverman, ``A robust method to extract talker azimuth
  orientation using a large-aperture microphone array,'' \emph{IEEE Trans.
  Audio, Speech, Lang. Process.}, vol.~18, no.~2, pp. 277--285, Feb. 2010.

\bibitem{Nakano2009}
A.~Y. Nakano, S.~Nakagawa, and K.~Yamamoto, ``Automatic estimation of position
  and orientation of an acoustic source by a microphone array network,''\,\emph{J.\,Acoust.\;Soc.\;Am.}, vol. 126, no.~6, pp. 3084--3094, Dec. 2009.

\bibitem{Takashima2011}
R.~Takashima, T.~Takiguchi, and Y.~Ariki, ``Single-channel head orientation
  estimation based on discrimination of acoustic transfer function,'' in
  \emph{Interspeech 2011}, Aug. 2011.

\bibitem{Takashima2012}
------, ``Estimation of talker's head orientation based on discrimination of
  the shape of cross-power spectrum phase coefficients,'' in \emph{Interspeech
  2012}, Sep. 2012.

\bibitem{Halkosaari2005}
T.~Halkosaari, M.~Vaalgamaa, and M.~Karjalainen, ``Directivity of artificial
  and human speech,'' \emph{J. Audio Eng. Soc.}, vol.~53, no. 7/8, pp.
  620--631, Jul. 2005.

\bibitem{Bellows2019a}
S.~D. Bellows and T.~W. Leishman, ``High-resolution analysis of the directivity
  factor and directivity index functions of human speech,'' in \emph{Audio Eng.
  Soc. Conv.}, Mar. 2019, pp. 1--10.

\bibitem{Poerschmann2020}
C.~P\"orschmann and J.~M. Arend, ``Analyzing the directivity patterns of human
  speakers,'' in \emph{Fortschritte der Akustik -- DAGA}, Mar. 2020, pp.
  1141--1144.

\bibitem{Chalker1985}
D.~Chalker and D.~Mackerras, ``Models for representing the acoustic radiation
  impedance of the mouth,'' \emph{IEEE Trans. Acoust. Speech. Signal Process.},
  vol.~33, no.~6, pp. 1606--1609, Dec. 1985.

\bibitem{Huopaniemi1999}
J.~Huopaniemi, K.~Kettunen, and J.~Rahkonen, ``Measurement and modeling
  techniques for directional sound radiation from the mouth,'' in \emph{Proc.
  1999 {IEEE} Workshop Applications of Signal Processing to Audio and Acoustics
  ({WASPAA})}, 1999.

\bibitem{Fischer2019}
G.~Fischer, C.~Schneiderwind, and A.~Neidhardt, ``Comparing the directivity of
  a mouth simulator and a simple physical model,'' in
  \emph{45\textsuperscript{th} Annual Meeting on Acoustics (DAGA)}, Mar. 2019.

\bibitem{Williams1999}
E.~G. Williams, ``Chapter 6 - spherical waves,'' in \emph{Fourier Acoustics},
  E.~G. Williams, Ed.\hskip 1em plus 0.5em minus 0.4em\relax London: Academic
  Press, 1999, pp. 183--234.

\bibitem{Graetzer2022}
S.~Graetzer, M.~A. Akeroyd, J.~Barker, T.~J. Cox, J.~F. Culling, G.~Naylor,
  E.~Porter, and R.~Viveros-Mu{\~{n}}oz, ``Dataset of british english speech
  recordings for psychoacoustics and speech processing research: The clarity
  speech corpus,'' \emph{Data in Brief}, vol.~41, p. 107951, Apr. 2022.

\bibitem{Habets2007}
E.~A.~P. Habets and S.~Gannot, ``Generating sensor signals in isotropic noise
  fields,'' \emph{J. Acoust. Soc. Am.}, vol. 122, no.~6, pp. 3464--3470, Dec.
  2007.

\end{thebibliography}
